# Dynamic tomography of the spin-orbit coupling in nonlinear optics

Hai-Jun Wu,[1] Zhi-Yuan Zhou,[1,2] Wei Gao,[1] Bao-Sen Shi,[1,2] and Zhi-Han Zhu[1,*]

[1]*Wang Da-Heng Collaborative Innovation Center for Quantum manipulation & Control, Harbin University of Science and Technology, Harbin 150080, China*

[2]*CAS Key Laboratory of Quantum Information, University of Science and Technology of China, Hefei, Anhui 230026, China*

*\*e-mail:* zhuzhihan@hrbust.edu.cn

Spin-orbit coupling (SOC) light fields with spatially inhomogeneous polarization have attracted increasing research interest within the optical community. In particular, owing to their spin-dependent phase and spatial structures, many nonlinear optical phenomena which we have been familiar with up to now are being re-examined, hence a revival of research in nonlinear optics. To fully investigate this topic, knowledge on how the topological structure of the light field evolves is necessary, but, as yet, there are few studies that address the structural evolution of the light field. Here, for the first time, we present a universal approach for theoretical tomographic treatment of the structural evolution of SOC light in nonlinear optics processes. Based on a Gedanken vector second harmonic generation, a fine-grained slice of evolving SOC light in a nonlinear interaction and the following diffraction propagation are studied theoretically and verified experimentally, and which at the same time reveal several interesting phenomena. The approach provides a useful tool for enhancing our capability to obtain a more nuanced understanding of vector nonlinear optics, and sets a foundation for further broad-based studies in nonlinear systems.

## 1. Introduction

Light fields, as a class of electromagnetic waves obeying full-vector Maxwell theory, manifest their vector nature via their states of polarization (SOPs), and the fields are broadly named vector light in the case that the SOPs are spatially inhomogeneous [1,2]. In recent years vector light with a custom spatial polarization structure plays an underpinning role in the science of structured light which has become one of today's most active and rapidly expanding fields of photonics [3,4]. From a historical perspective, research on vector light, or rather the special category named cylindrical vector modes or well-known waveguide modes [2,5], can be traced back some 60 years ago (only a decade after Townes *et al*. invented the laser), a group of scientists began research on laser transverse electric mode [6,7]. Thereafter, many researchers focused on these laser counterparts of Maxwell equation's vector solutions. Until 1992, Allen *et al.* published their milestone paper that opened up research on light's orbital angular momentum (OAM) [8-10], the work on finding the connection between vector light and OAM leads to the emerging of word 'vector vortex'. More recently, due to the advance in micro-nano photonics and the extending of quantum optics from single photon to classical light [11-13], two underlying mechanisms behind the spatially variant SOPs, i.e., spin-orbit coupling (SOC) and geometric phase (or Pancharatnam–Berry phase) [14-17], have been gradually revealed in the last decade. The emergence of the above two concepts leads to a new paradigm for modern optics impacting several fields. In particular, first, it enables the analysis and manipulation of structured light as SOC entangled states [18-22] and, second, it underlies spin-dependent shaping and control of structured light via geometric phase manipulation [23-28].

The advances in structured light have also been responsible for revisions regarding our understandings of nonlinear optics [29-39]. The quintessential attribute of vector light, SOC, on the one hand, gives rise to significant spin-dependent phase and intensity properties of light and, hence, can dramatically tailor optical nonlinear interactions, while, on the other hand, the SOC mediated vector nonlinear interactions provide a feasible interface for shaping and controlling structured light as well. Besides, the transition and evolution of the topological structure in vector nonlinear interactions offer new insights into the physics underlying the transition of geometric phases in a SOC system. The central premise for investigations on the above issues is a good knowledge of nonlinear optics with vector light fields. To date, despite some recent works that have focused on this topic [40-44], a demonstration on how best to efficiently and conveniently analyze the structure evolution of vector light during nonlinear optics processes has yet to be proposed. To address this, we present a universal theoretical approach that facilitates dynamic tomography of the fine structure of the SOC light in nonlinear optics processes. We have achieved this by calculating the vector paraxial path integral with nonlinear beating fields as pupil functions, so that the obtained vector wavefunctions can describe the full structure of vector light in a nonlinear interaction and following diffraction propagation.

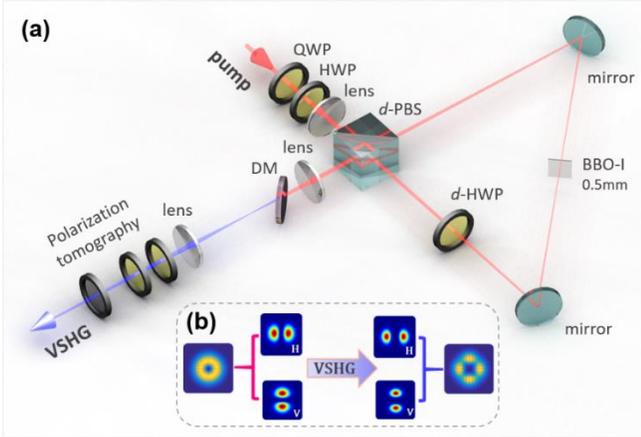

FIG. 1. Schematic of experimental setup (a) and the principle of VSHG (b). The VSHG is realized by a polarizing Sagnac loop that comprises a dual-wavelength polarizing beam splitter ($d$-PBS), a dual-wavelength half-wave plate ($d$-HWP) and a 0.5 mm type-I BBO. The vector pump light is shaped (complex amplitude modulation) 795nm 150 fs pulse light, and a pair of waveplates (QWP and HWP) is employed for modulating it to a desired vector mode. At output port, a 4f imaging system is employed for observing VSHG light's evolution from the pupil plane to the far field. At last, pair of waveplates and a polarizer is used for polarization tomography.

The specific nonlinear interaction that we use for this demonstration is the accessible Second Harmonic Generation (SHG). It should be noted, however, that typical second order nonlinear optical processes usually are linear polarization dependent. More specifically, for vector light, only one polarization part can experience nonlinear conversion in type-I SHG; and in type-II SHG, it is actually a Sum Frequency Generation between the two orthogonal linear polarized parts. Therefore, to conveniently demonstrate this approach, we propose a Gedanken second order nonlinear optical process, called vector SHG (VSHG), where the two orthogonal polarization parts of the input vector light can be simultaneously converted into SHG light, as shown in Fig. 1. This proposal can be realized experimentally by the optical setup of Fig. 1, in which a self-locking nonlinear interferometer with type-I crystal, first described by Shi and Tomita [45,46], is used. In the following, we first present the theoretical approach; then, with theoretical tomography we show how the structure of vector light evolves in the VSHG, with cylindrical vector and Full-Poincaré light as specific models, respectively; finally, the tomography is verified by experimentation.

## 2. Theory

The term vector light field generally refers to structured light in the case where the SAM and OAM are non-separable with each other, i.e., SOC entangled states. The light field can be described as a vector superposition of two Laguerre-Gaussian (LG) modes (see Eq. (A1) in Appendix) with mutually orthogonal polarizations

$$\hat{E}(r,\varphi,z) = \sqrt{\alpha}LG_{+\ell}(r,\varphi,z)\hat{\mathbf{e}}_+ + e^{i\theta}\sqrt{1-\alpha}LG_{-\ell}(r,\varphi,z)\hat{\mathbf{e}}_-, \quad (1)$$

where $LG_{+\ell}$ and $LG_{-\ell}$ denote the polarization dependent LG modes (with the radial index of LG modes $p=0$) corresponding to two orthogonal polarizations $\hat{\mathbf{e}}_+$ and $\hat{\mathbf{e}}_-$, respectively, where $e^{i\theta}$ is the intra-model phase. The mode weight coefficient $\alpha \in [0,1]$ represents the degree of 'entanglement', in which the light is a full vector for $\alpha=0.5$ and is scalar for $\alpha=0$ or $1$ [18], respectively. Here, we focus on the full vector light, i.e., $\alpha=0.5$, therefore, we can reformulate the SOC states into a simple Dirac notion

$$\left|\psi_{\mathbf{k}(\omega)}\right\rangle = \left(\left|\hat{\mathbf{e}}_+,\psi_\ell^+\right\rangle + e^{i\theta}\left|\hat{\mathbf{e}}_-,\psi_\ell^-\right\rangle\right)/\sqrt{2}, \quad (2)$$

where the spin-dependent spatial mode is given by $\left|\psi_\ell^+\right\rangle$ and $\left|\psi_\ell^-\right\rangle$, and $\mathbf{k}(\omega)$ is the dispersion relation representing the light field's wave vector. These SOC states can be divided into different categories depending on the topological phase of the spatial modes. Among them, the two most commonly encountered categories are: (1) the cylindrical vector mode, for two complementary topological phases $\ell_+ = -\ell_-$, whose spatially variant SOPs map to a certain point on the equator of the corresponding higher-order Poincaré sphere (HOPS), for more details see Ref.46 and 47; (2) the Full-Poincaré modes, for the case $\ell_+ \neq -\ell_-$, whose transverse SOPs can fully cover at least one surface of a scalar polarization Poincaré sphere [1,48]. Below, the VSHG specific to these two special categories are discussed.

As a well-known nonlinear optics phenomenon, SHG can be described as a pump (signal) light $E_{p(s)} = A_{p(s)} \exp(ik(\omega)z)$, beating with itself in a nonlinear crystal and exciting a SHG light $E_{SHG} = A_{SHG} \exp(ik(2\omega)z)$ in the same direction, and the corresponding nonlinear wave equation can be expressed as

$$\partial E_{SHG} / \partial z = -iT\kappa E_p E_s \exp[i\Delta k z], \quad (3)$$

where $\kappa$ is the coupling coefficient, $\Delta k$ is the degree of phase-mismatching, and T ( $T = 0.5$ ) is the degenerate coefficient for SHG. By calculating this three-wave coupling equation with the finite-element method [39], one can numerically analyze the target parameter in the dynamic process, especially for energy flux and phase transitions; however, it is difficult to use for tomographic fine structure evolution. To achieve a more effective tomography, here we employ a paraxial path integral method. Notice that the term $E_p E_s$ in the right side of Eq. (3) describes a time-varying beating field, which drives the oscillation of the charge in nonlinear medium, or rather, the beating field acts as the diffraction source of the excited SHG field [50]. Therefore, the pupil (or aperture) function, which can be used to predict the diffraction propagation of the excited SHG field, at the generation plane $z_0 = 0$ can be expressed as

$$E_{pupil}(r_0, \varphi_0, 0) = E_p(r_0, \varphi_0, 0) E_s(r_0, \varphi_0, 0), \quad (4)$$

where $E_p = E_s$ for type-I SHG; and it is assumed that the nonlinear interaction finishes in a thin slice along the z-axis. Consider that the pupil field is coherently constructed by LG modes which are eigen modes of the paraxial wave equation (PWE) carrying constant energy. Therefore, we adopt the paraxial Collins propagator to derive the scalar wavefunction of the excited SHG field upon diffraction propagation [37,51], which is given by

$$E_{SHG}(r, \varphi, z) = \frac{i}{\lambda z} \exp(-ikz) \int r_0 dr_0 \int d\varphi_0 E_{pupil}(r_0, \varphi_0, 0)$$
$$\exp\left\{-\frac{ik}{2z} \times [r_0^2 - 2rr_0 \cos(\varphi - \varphi_0) + r^2]\right\}. \quad (5)$$

For a given pupil function input, this propagator can give an analytical expression for the corresponding SHG field. It should be noted that although the beating fields are constructed by PWE eigen modes whose beam profile are constant upon propagation; the pupil functions describe complex (both intensity and phase) modulated PWE modes. Hence, the beam profile of SHG fields may not propagating constant, especially for the case of using structured light as fundamental frequency fields. In the case of VSHG, according to Eq. (2), the pupil function shall become of vector form and can be expressed with respect to the orthogonal circular-polarization basis

$$\left|\Psi_{\mathbf{k}(2\omega)}\right\rangle_{pupil}^{L,R} = [\sqrt{\beta} \langle \hat{\mathbf{e}}_L | \psi \rangle^2 \hat{\mathbf{e}}_L + e^{i\eta} e^{i\theta} \sqrt{1-\beta} \langle \hat{\mathbf{e}}_R | \psi \rangle^2 \hat{\mathbf{e}}_R], \quad (6)$$
$$= [\sqrt{\beta} |\psi_L^\ell\rangle^2 \hat{\mathbf{e}}_L + e^{i\theta} \sqrt{1-\beta} |\psi_R^\ell\rangle^2 \hat{\mathbf{e}}_R]$$

or the orthogonal linear-polarization basis via the relations $(\hat{\mathbf{e}}_H - i\hat{\mathbf{e}}_V)/\sqrt{2} = \hat{\mathbf{e}}_L$ and $(\hat{\mathbf{e}}_H + i\hat{\mathbf{e}}_V)/\sqrt{2} = \hat{\mathbf{e}}_R$

$$\left|\Psi_{\mathbf{k}(2\omega)}\right\rangle_{pupil}^{H,V} = [\sqrt{\beta} \langle \hat{\mathbf{e}}_H | \psi \rangle^2 \hat{\mathbf{e}}_H + e^{i\eta} e^{i\theta} \sqrt{1-\beta} \langle \hat{\mathbf{e}}_V | \psi \rangle^2 \hat{\mathbf{e}}_V],$$
$$= [\sqrt{\beta}(|\psi_L^\ell\rangle + e^{i\theta}|\psi_R^\ell\rangle)^2 \hat{\mathbf{e}}_H + e^{i\eta}\sqrt{1-\beta}(|\psi_L^\ell\rangle - e^{i\theta}|\psi_R^\ell\rangle)^2 \hat{\mathbf{e}}_V]$$
$$(7)$$

where the mode weight coefficient $\beta$ depends on the relative SHG efficiency between the different spatial modes, and $e^{i\eta}$ is the intra-polarization phase factor added by the nonlinear system; for simplicity and without loss of generality, it is assumed that $e^{i\eta} = 1$. Besides, notice that the efficiency of nonlinear conversion depends on power density, thus the SHG involving structured pump will lead to a spatial mode dependent $\beta$ (or spatial mode dependent beam size and overlap between fundamental frequency fields). In this paper, $\beta$ for all example fields are equal or approximately equal to 0.5, and a more detailed analysis is the subject of a future paper. Combining Eqs. (5), (6), and (7), we can obtain a vector wavefunction that can fully describe how the structure of the SHG light evolves upon propagation, and which can be expressed as

$$\hat{E}_{SHG}(r,\varphi,z) = E_{SHG}^{+}(r,\varphi,z)\hat{\mathbf{e}}_{+} + E_{SHG}^{-}(r,\varphi,z)\hat{\mathbf{e}}_{-}. \quad (8)$$

After getting this vector wavefunction, we can visualize the full transverse structure of the VSHG light at a given propagation distance. Note, the polarization structure here is obtained through the higher-order Stokes parameters, first introduced by Milione et al. [47,48], in which the functions of polarization ellipticity $\chi$ and orientation $\phi$, which are dependent on the transverse point $\{r,\varphi\}$, are given by

$$\chi = \frac{1}{2}\sin^{-1}\left(\frac{S_3^\ell}{S_0^\ell}\right) \text{ and } \phi = \frac{1}{2}\tan^{-1}\left(\frac{S_2^\ell}{S_1^\ell}\right), \quad (9)$$

where $S_{0-3}^\ell$ are the corresponding $\ell$-dependent higher-order Stokes parameters.

### 3. Dynamic tomography for cylindrical vector modes

Now, we start from cylindrical vector modes with an intra-model phase $e^{i\theta}=1$, or rather radial polarized light, to demonstrate the dynamic tomography for the VSHG. Thus, the Dirac notion shown in Eq. (3) can be simplified as $(|\hat{\mathbf{e}}_L,LG_{+\ell}\rangle+|\hat{\mathbf{e}}_R,LG_{-\ell}\rangle)/\sqrt{2}$. Then, two cases should be considered separately: first, the VSHG takes place within two opposite circular-polarization basis; second the VSHG occurs within two orthogonal linear-polarization basis. For the first case, according to Eq. (6), the pupil function can be expressed as

$$\left|\Psi_{CV}\right\rangle_{pupil}^{L,R} = \left(\left|\hat{\mathbf{e}}_L,LG_{+\ell}^{\;2}\right\rangle + \left|\hat{\mathbf{e}}_R,LG_{-\ell}^{\;2}\right\rangle\right)/\sqrt{2}. \quad (10)$$

The expression depicts a topological phase $\exp(i\ell\varphi)$ doubling in the OAM subspace of the SOC state; in consequence, the VSHG transition leads to a 'jump' between two cylindrical vector mode HOPS, i.e., $|\hat{\mathbf{e}}_L,+\ell\rangle+|\hat{\mathbf{e}}_R,-\ell\rangle \rightarrow |\hat{\mathbf{e}}_L,+2\ell\rangle+|\hat{\mathbf{e}}_R,-2\ell\rangle$, as shown in Fig. 2(a). After substituting Eq. (10) into Eq. (5), we get the corresponding vector wavefunctions of the VSHG fields, and Fig. 2(b) shows the simulated dynamic tomography. The simulation presents an intuitional result, specifically, the structures of VSHG light are very close to their PWE eigen mode analogs, i.e., LG-mode based SOC states, and are stable upon propagation. The only difference is that the ring of the VSHG light are relatively narrow at the generation plane due to the intensity modulation (beam shrinking in radial dimension) of nonlinear interaction, namely, this VSHG light can be regarded as quasi-eigen states of PWE. This intuitional scenario, however, will alter dramatically in the second case which we will consider next.

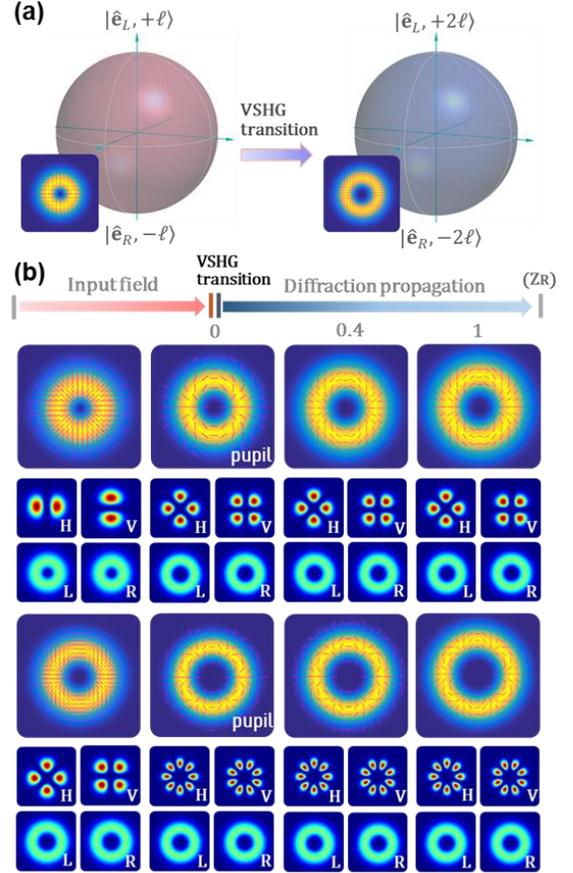

FIG. 2. (a) Schematic of L- and R-polarized VSHG transition for cylindrical vector modes. (b) Nonlinear dynamic tomography for cylindrical vector modes ($\ell=1,2$) in L- and R-polarized VSHG.

For the second case, after reforming the cylindrical vector mode within the linear-polarization $\hat{\mathbf{e}}_H$ and $\hat{\mathbf{e}}_V$ basis, the vector pupil function, according to Eq. (6), can be expressed as

$$\left|\Psi_{CV}\right\rangle_{pupil}^{H,V} = \left(\left|\hat{\mathbf{e}}_H,(|LG_{+\ell}\rangle+|LG_{-\ell}\rangle)^2\right\rangle + \left|\hat{\mathbf{e}}_V,(|LG_{+\ell}\rangle-|LG_{-\ell}\rangle)^2\right\rangle\right)/\sqrt{8}, \quad (11)$$

where for the case $\ell=1$ it can be represented with Hermit-Gauss modes as $(|\hat{\mathbf{e}}_H,HG_{10}^{\;2}\rangle+|\hat{\mathbf{e}}_V,HG_{01}^{\;2}\rangle)/\sqrt{2}$. Such a pupil function indicates that the VSHG transition at this time is much stronger than that in the former case, leading to a significant 'jump' from a HOPS of the standard cylindrical vector mode to a more complex one, i.e., $\lceil(|+2\ell\rangle+|-2\ell\rangle+2|0\rangle)\hat{\mathbf{e}}_H+(|+2\ell\rangle+|-2\ell\rangle-2|0\rangle)\hat{\mathbf{e}}_V\rceil/\sqrt{8}$, which we are unfamiliar with, as shown in Fig. 3(a). The dynamic

tomography for $\ell = 1, 2$ is shown in Fig. 3(b). The tomography shows that, compared to the PWE-eigen-mode analogs, the structure evolution of VSHG fields are unusual, interesting, and worth studying.

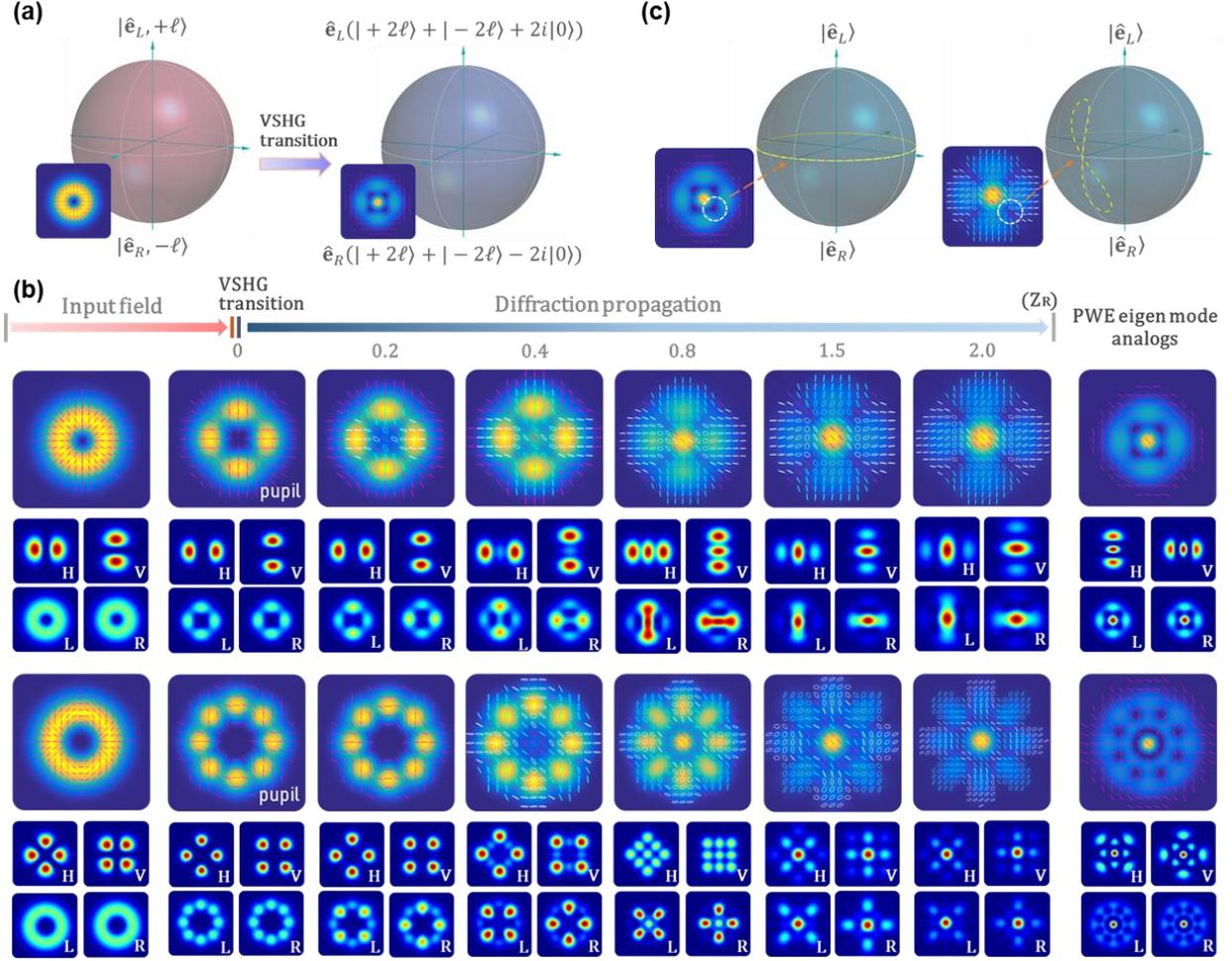

FIG. 3. (a) Schematic of H- and V-polarized VSHG transition for cylindrical vector modes. (b) Nonlinear dynamic tomography for cylindrical vector modes ( $\ell = 1, 2$ ) in H- and V-polarized VSHG, where the green and white circles on the transverse planes represent L- and R-circular polarizations, respectively, and the rightmost column is PWE-eigen-mode analogs for comparison. (c) Structure difference between the VSHG light and its PWE-eigen-mode analog, where the SOPs round the white dotted circles maps to the yellow dotted line on the PS.

First, despite the VSHG light carries the same SOC structures as their PWE-eigen-mode analogs, they are not eigen modes of free-space PWE as discussed in section 2, and, therefore, they exhibit a strange unstable structure upon propagation. Specifically, the VSHG light originates from the beating field of the input pump, as shown in Eq. (11) and Fig. 3(b) at $Z_R = 0$, and on this basis the topological phase transition $(|+\ell\rangle \pm |-\ell\rangle)/\sqrt{2} \rightarrow (|+2\ell\rangle + |-2\ell\rangle \pm 2|0\rangle)/\sqrt{4}$, which occurred in orthogonal linear polarizations, lead to a drastic profile evolution in the following diffraction propagation. As a consequence, the structures, that is, the intensity, phase, and polarization of the VSHG light fields, whatever in the near or far field, are very different from their PWE-eigen-mode analogs.

Second, a notable difference is that elliptical polarization can be found in the far field of the VSHG light, but not for the PWE eigen modes and the near field of VSHG light. This phenomenon can be attributed to the azimuthal position squeezing in OAM space, or, in other words, broadening of the OAM spectrum [52], which is induced by azimuthal position dependent intensity modulation nonlinear interaction. To be specific, the VSHG within $\hat{\mathbf{e}}_H$ and $\hat{\mathbf{e}}_V$ leads to an inhomogeneous intensity distribution, or a beam shrinking effect, in azimuthal dimension at the generation plane as shown in Fig. 3(b) at $Z_R = 0$. Here, a specific polarization singular point on the transvers plane was chosen as an entry point to highlight the interesting difference revealed by the tomography. As shown in Fig. 3(c), in the PWE eigen mode the chosen singular point is surrounded by an 'L-line' that maps into the equator on the polarization Poincaré sphere; in contrast, the surrounding polarization states in the VSHG case turn into an equator-centered "8-like" loop that is across two hemispheres. Such interesting azimuthal position squeezing will be the subject of future studies on optical parametric amplification systems with specialized theoretical and experimental tools [18,19].

In addition to above salient points, one may find an implied physical insight for an old physical puzzle — Loschmidt's or the reversibility paradox — that is, how to obtain an irreversible process from time-symmetric dynamics [53]? It is well known that the wave equations are time-reversal invariance, in consequence, knowing the information of a wave reaching a 2D surface can be used to recover the wave dynamics occurred in the past. In nonlinear wave dynamics, time-reversal symmetry indicates the reversibility of nonlinear conversion, for example, for a given SHG, there should be a down-frequency generation (DFG) process can act as its time-reversal dynamics. However, in contrast to the first case shown in Fig. (2), here the VSHG transition leads to non-reciprocal wave dynamics. That is to say one cannot infer the SOC state of pump from the VSHG light via a time-reversal DFG, i.e., obviously, according to Eq. (11) $\hat{E}_{SHG}\hat{E}_p^* \neq \hat{E}_p$. Recently, from a quantum perspective, we have known that the collapse of wavefunctions in projection measurement (due to entropy increase or information loss) can break the time-reversal symmetry of wave dynamics [54]. Unlike time symmetry breaking induced by the wavefunction collapse, the non-reciprocal wave dynamics revealed in this study may originate from the topological phase transition $|+\ell\rangle \pm |-\ell\rangle \to |+2\ell\rangle + |-2\ell\rangle \pm 2|0\rangle$, which makes the OAM state escape from SU(2) space; more detailed analysis will be the subject of future work.

## 4. Dynamic tomography for Full-Poincaré modes

In this section, we continue to analyze the Full-Poincaré modes with the most common form, i.e., $(|\hat{\mathbf{e}}_L, 0\rangle + |\hat{\mathbf{e}}_R, -\ell\rangle)/\sqrt{2}$. Similar to the analysis above, the VSHG have been considered to take place in orthogonal circular or linear polarizations, respectively, and the pupil functions according to the above discussion can be expressed as

$$|\Psi_{FP}\rangle_{pupil}^{L,R} = \left(|\hat{\mathbf{e}}_L, LG_0^2\rangle + |\hat{\mathbf{e}}_R, LG_{-\ell}^2\rangle\right)/\sqrt{2} \qquad (12)$$

and

$$|\Psi_{FP}\rangle_{pupil}^{H,V} = \left(|\hat{\mathbf{e}}_H, (|LG_0\rangle+|LG_{-\ell}\rangle)^2\rangle + |\hat{\mathbf{e}}_V, (|LG_0\rangle-|LG_{-\ell}\rangle)^2\rangle\right)/\sqrt{2}, \qquad (13)$$

where, like similar cases in cylindrical vector modes, Eq. (12) describes an intuitionally $\ell$-doubling jump between diffrenent order Full-Poincaré mode HOPS, whereas Eq. (13) describes a significant jump that leads to non-reciprocal wave dynamics, $(|0\rangle+|\ell\rangle)\hat{\mathbf{e}}_H + (|0\rangle-|\ell\rangle)\hat{\mathbf{e}}_V \to (|0\rangle+2|\ell\rangle+|+2\ell\rangle)\hat{\mathbf{e}}_H + (|0\rangle-2|\ell\rangle+|+2\ell\rangle)\hat{\mathbf{e}}_V$, as shown in Fig. 4(a). In this section we focus on the second case, the dynamic tomography for the first case can be found in Fig. B1 in Appendix B.

Figure 4(b) shows the simulated dynamic tomography for $\ell = 1, 2$. It can be seen that, unlike the propagation variant structure shown in the cylindrical vector modes, here the transverse structures are propagation constant. This is because the pupil functions' profile $(|LG_0\rangle \pm |LG_{-\ell}\rangle)^2$ are similar with the SHG fields' PWE-eigen-mode analogs $(|0\rangle + 2|\ell\rangle + |+2\ell\rangle)$ at $z_0 = 0$, and the slight difference in polarization comes from the beam shrinking effect of SHG. Moreover, compared with cases in the cylindrical vector modes, another difference here is that the VSHG light's profiles experience a clockwise rotation upon diffraction propagation that can achieve a $\pi/2$ total rotation angle at the far field. The 3D curve in Fig. 4(c) shows the simulated rotation of polarization singularity structure based on the tomography. Such regular rotation results from the fact that, for LG modes, the speed of Gouy phase accumulation are $\ell$-dependent, thus, an extra intra-mode phase between different LG modes will appear during the propagation. To be specific, according to Eq. (A1), the rotation of the inference profile $\langle \ell_1 | \ell_2 \rangle$ can be expressed as $\exp[i(|\ell_1|-|\ell_2|)\tan^{-1}(z/z_R)]$, thus the total phase variation at a given transverse point is $(|\ell_1|-|\ell_2|)\pi/2$ that corresponds to a $\pi/2$ total rotation for the light profile.

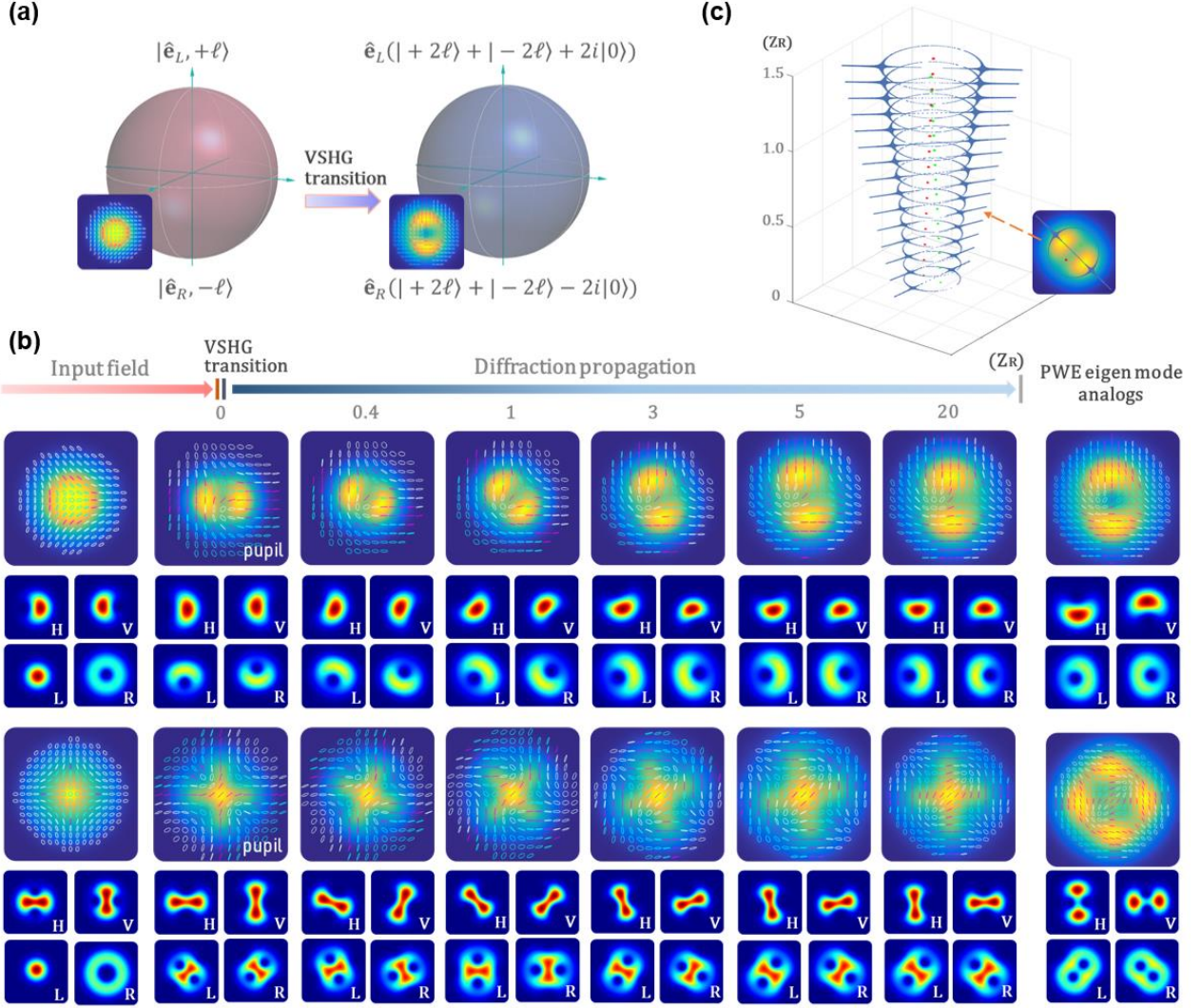

FIG. 4. (a) Schematic of H- and V-polarized VSHG transition for Full-Poincaré modes. (b) Nonlinear dynamic tomography for Full-Poincaré modes ($\ell = 1, 2$) in H- and V-polarized VSHG, where the green and white circles on the transverse planes represent L- and R-circular polarizations, respectively, and the rightmost column is PWE-eigen-mode analogs for comparison. (c) Polarization rotation of the VSHG light ($\ell = 1$) upon propagation, where blue line depicts 'L-line', green and red spot depict 'C-points'.

## 5. Experimental results

To demonstrate experimentally the accuracy of the tomography, a self-locking Sagnac Interferometer with type-I BBO crystal was used for realizing the VSHG, as shown in Fig. 1, where a 4$f$ imaging system was employed to observe the dynamic evolution of VSHG light from the pupil to the far field. However, it should be noted that the most commonly used vortex light obtained via phase-only manipulation, such as SLM and q-plate, is a kind of hypergeometric Gaussian mode [55], which contains undesired radial parameters or rather can be represented as a LG superposed mode with the same $\ell$ and different $p$. Therefore, for verification of the accuracy of the theory, complex amplitude modulation was used to generate the SOC light based on pure LG modes [56]. In this section, we have focused on verifying the interest prediction that will be observed in linear-polarization based VSHG. In the experiment, the structure evolution of VSHG light generated from the cylindrical vector modes was first observed, including the intensity profile from the pupil plane to the far field and polarization tomography in the far field. Figure 5 shows the observed results (left column) with $\ell = 1 \sim 5$; for comparison purposes, simulated data are also presented in the right column. It can be seen that the experimental results agree well with the theoretical tomography. Then, we observed the structure evolution of VSHG light generated from the Full-Poincaré modes. Again, the observed structure evolution and propagation rotation shown in Fig. 6 confirm the accuracy of the theoretical approach.

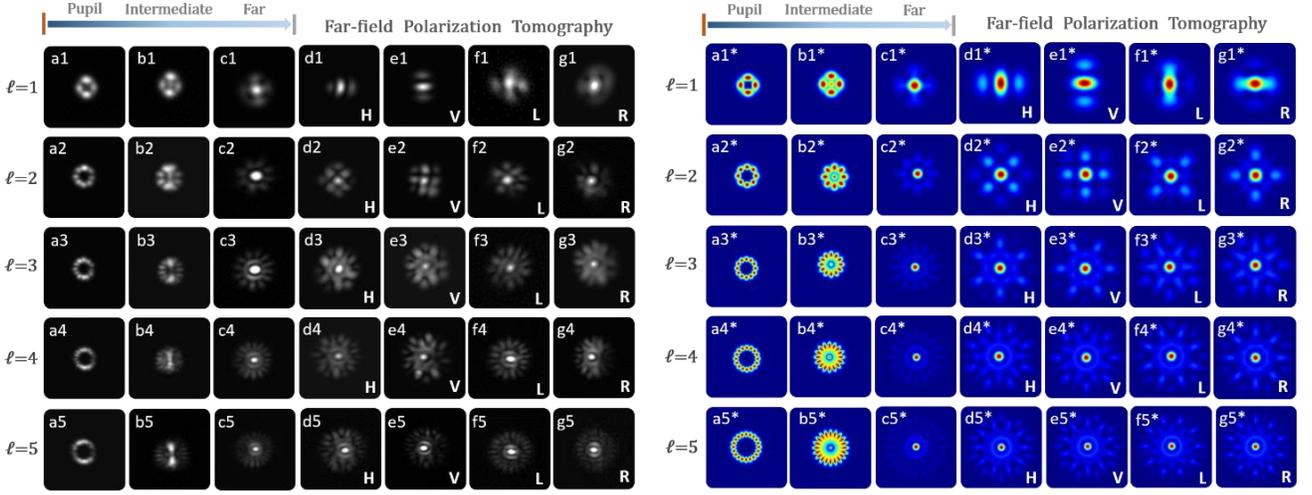

FIG. 5. Observed Nonlinear dynamic evolution of cylindrical vector modes in H- and V-polarized VSHG (left column), and the simulated observables for comparison (right column).

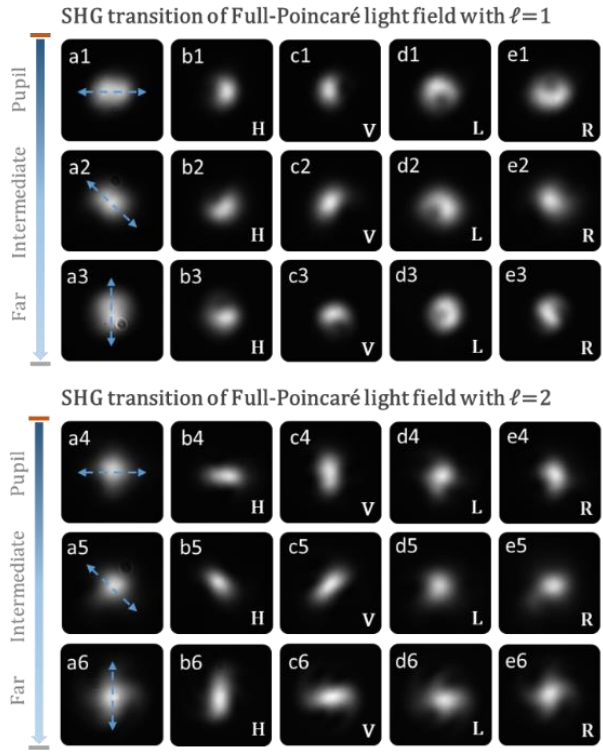

FIG. 6. Observed Nonlinear dynamic evolution of Full-Poincaré modes ($\ell = 1, 2$) in H- and V-polarized VSHG.

## 6. Conclusion

The dynamic tomography presented shows a universal theoretical approach that has enhanced our capability to study nonlinear optics with structured light, and, as a consequence, the vectorial optical nonlinear processes can be readily analyzed to an unprecedented degree. Particularly, the specific example used in this demonstration, VSHG, owing to such precise tomography, unveils a fine-grained evolution slice of SOC light in the VSHG and in the following diffraction propagation (induced by intensity and phase modulation of nonlinear interaction), revealing several potential interesting phenomena and physical insights. In a word, this tomography method can play an important role in providing a more nuanced understanding of vector nonlinear interactions in broadly-based systems, such as stimulated Raman/Brillouin scattering and light-atom interactions.

## Funding

This work was supported by National Natural Science Foundation of China (Grant Nos. 11574065, 11604322, and 61525504)

## Appendix A

The wavefunction of Laguerre-Gaussian (LG) mode in cylindrical coordinates $\{r, \varphi, z\}$ can be expressed as [9]

$$LG_\ell^p(r,\varphi,z) = \sqrt{\frac{2p!}{\pi(|\ell|+p)!}} \frac{1}{w(z)} \left(\frac{\sqrt{2}r}{w(z)}\right)^{|\ell|}$$
$$L_p^{|\ell|}\left(\frac{2r^2}{w^2(z)}\right) \exp\left(-\frac{r^2}{w^2(z)}\right) \exp(-i\Phi(r,\varphi,z))$$

$$\Phi(r,\varphi,z) = kz + \omega r^2/2c\,R(z) + \ell\varphi - (2p-|\ell|+1)\tan^{-1}(z/z_R)$$
,(A1)

where $\ell$ is the topological charge giving an OAM of $\ell\hbar$ per photon, $p$ is the radial index of $LG$ modes, $L_p^{|\ell|}$ is the Laguerre polynomial, R($z$) is the curvature radius of the wavefront, $z_R$ is the Rayleigh length for a given beam waist $w_0$, $w(z) = w_0(1+z^2/z_R^2)^{-1}$ and $(2p-|\ell|+1)\tan^{-1}(z/z_R)$ describes the beam expanding and the Gouy phase accumulated during the diffraction propagation, respectively.

## Appendix B

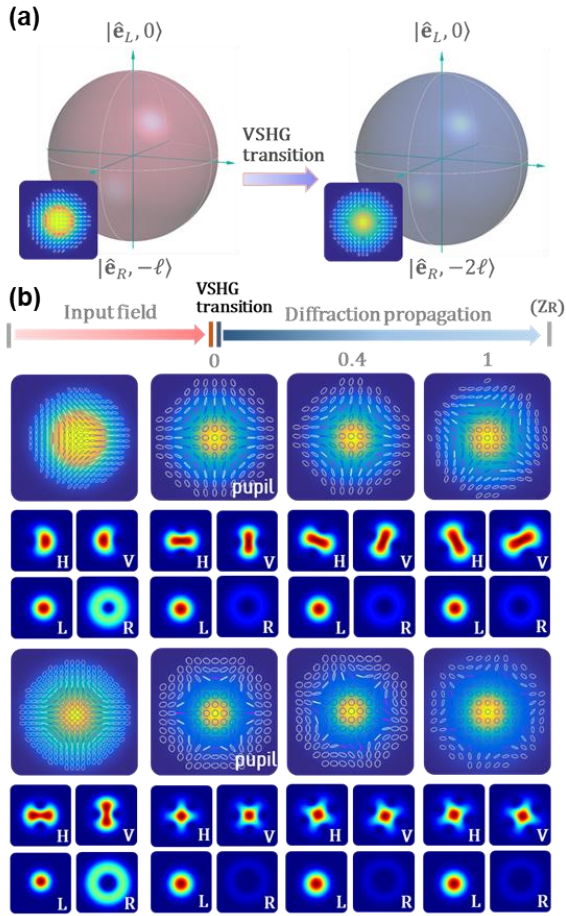

FIG. B1. (a) Schematic of L- and R-polarized VSHG transition for Full-Poincaré modes. (b) Nonlinear dynamic for Full-Poincaré modes ($\ell = 1, 2$) in L- and R- polarized VSHG.